# AlfaMC: a fast alpha particle transport Monte Carlo code

Luis Peralta[1,2], Alina Louro[1]

[1] Faculdade de Ciências da Universidade de Lisboa

[2] Laboratório de Instrumentação e Física Experimental de Partículas

**Abstract**

AlfaMC is a Monte Carlo simulation code for the transport of alpha particles. The code is based on the Continuous Slowing Down Approximation and uses the NIST/ASTAR stopping-power database. The code uses a powerful geometrical package allowing the coding of complex geometries. A flexible histogramming package is used which greatly easies the scoring of results. The code is tailored for microdosimetric applications where speed is a key factor. Comparison with the SRIM code is made for transmitted energy in thin layers and range for air, mylar, aluminum and gold. The general agreement between the two codes is good for beam energies between 1 and 12 MeV. The code is open-source and released under the General Public Licence.

**1. Introduction**

Alpha particles are highly ionizing and have low penetration in matter. Their energy ranges from 1.830 MeV ($^{144}$Nd) to 11.740 MeV ($^{266m}$Mt) (Table, 2004) although most common alpha sources will emit particles in the 4 to 9 MeV range (Table, 2004). In condensed matter their range is typically below 100 µm, while in air at STP they can reach a few centimeters (Berger, 2011; Turner, 2007). The energy transfer from an alpha particle to atomic electron in a single collision is small. For non-relativistic particles of relative velocity β and $\gamma = 1/\sqrt{1-\beta^2}$ the maximum energy transfer $T_{max}$ to an electron of mass $m_e$ is approximately given by (Beringer, 2012) $T_{max} \approx 2m_e c^2 \beta^2 \gamma^2$ where c is the speed of light. For an alpha particle of 10 MeV this results in a maximum energy transfer of only 5.4 keV. These electrons will lose their energy in the vicinity of the interaction point with the alpha particle and their transport can be neglected by the Monte Carlo code. Due to the alpha particle high mass, the emission of bremsstrahlung radiation at these energies is completely negligible. For these reasons the Continuous Slowing Down Approximation (CSDA) (Turner, 2007) which assumes that a charged particle loses energy continuously along its path at the linear rate given by the instantaneous stopping power is an adequate approximation at this energy range.

General purpose Monte Carlo programs as MCNPX, GEANT3, GEANT4 or FLUKA (Agostinelli, 2003; Allison, 2006; GEANT, 1993; MCNPX, 2012; FLUKA, 2012) can simulate the transport and energy loss of alpha particles in matter. They can simulate complex geometries, but their speed is in general low. The state-of-the-art SRIM program (SRIM, 2008; Ziegler, 1985) on the other hand is relatively fast but allows only simple slab geometry. To overcome some of these problems several application specific codes have been developed. Unfortunately, for most of them there is a limited access to the code or even to documentation describing the underlying models.

There is thus the need for an open-source fast Monte Carlo package capable of dealing with complex geometries. The AlfaMC package was developed to fill this gap. The package is able to simulate complex geometrical bodies and spatially-distributed alpha particle sources. It has a sophisticated scoring and histogramming set of routines. The package uses the National Institute of Standards and Technology (NIST) ASTAR database where stopping powers for 74 materials (26 elements and 48 compounds and mixtures) are available. Stopping power data for new materials based on this set can be generated using the Alfamaterial.f program. The alpha particle transport is based on the Continuous Slowing Down Approximation. Gaussian or Landau distributed energy straggling is performed. A simple Fermi small-angle multiple scattering model is adopted.

## 2. AlfaMC physical model
## 2.1 Stopping-power computation

Charged particles interact with the electrons and nuclei of the medium primarily through the Coulomb electric force . For alpha particles most of these interactions individually transfer only minute fractions of the incident particle's kinetic energy. It is a convenient approximation to regard the alpha particle as losing its kinetic energy gradually and without hard collisions with the atomic electrons in a process referred to as the "Continuous Slowing-Down Approximation" (CSDA).

Based on the CSDA the AlfaMC code assumes that average energy loss per unit length is given by the unrestricted stopping power of the alpha particle. The NIST ASTAR database supplies separate values for the electronic and nuclear stopping power. The AlfaMC adds both values to obtain the total stopping power value. The original ASTAR tables present stopping power values for 122 energy values between 0.001 MeV and 1000 MeV, distributed on a logarithmic grid. The AlfaMC code then computes the stopping power values S' on a linear energy grid of 0.001 MeV in the energy range of 0.001 MeV to 100 MeV. To obtain the interpolated S values a logarithmic interpolation of the ASTAR stopping power S values is made as follows: let E be an energy value between two E1 and E2 consecutive energy values form the original table with stopping power values S1 and S2. Then the stopping power S for E is obtained using the formula

$$\ln S = \frac{\ln S_2 - \ln S_1}{\ln E_2 - \ln E_1}(\ln E - \ln E_1) + \ln S_1 \qquad (1)$$

or rearranging the expression in terms of S

$$S = S_1 (E/E_1)^m \qquad (2)$$

where

$$m = \frac{\ln S_2 - \ln S_1}{\ln E_2 - \ln E_1} \quad . \tag{3}$$

For unevenly spaced grids and non-linear functions and the logarithmic interpolation will give a better approximation to the function value than the one that can be obtained with a linear interpolation. The use of precomputed lookup tables greatly improves the computation time, since the computation of stopping powers is one of the most performed operations. The precomputed S value depends on the energy and its extraction from the table is made using the energy related index **i=int(E*1000)** since the table energy interval is 1/1000 MeV.

**2.2 Energy straggling**

The total continuous energy loss of an alpha particle is a stochastic quantity with a distribution often called the straggling function. In the tracking model followed by AlfaMC no secondary particles are produced and the continuous energy loss is unrestricted. For thick slabs the energy loss distribution approaches a Gaussian (Leo, 1994) but for thin traversed slabs other distributions can be considered (Leo, 1994; GEANT, 1993; Tsoulfanidis, 1995). A cut parameter $\kappa$ can be defined to set the limit between the thin and thick slab as (GEANT, 1993; Leo, 1994) $\Delta E = \kappa T_{max}$ where $\Delta E$ is the energy deposited in the slab and $T_{max}$ the maximum energy transfer in a single collision given by (Beringer, 2012)

$$T_{max} = \frac{2 m_e c^2 \beta^2 \gamma^2}{1 + 2\gamma m_e/m_\alpha + m_e^2/m_\alpha^2} \tag{4}$$

where $m_\alpha$ is the alpha particle mass.

Usually it is assumed that for $\kappa > 10$ the Gaussian model is a good approximation, while for the intermediate values $0.01 \leq \kappa < 10$ the straggling function follows a Vavilov distribution (Vavilov, 1957; Schorr, 1974). For lower $\kappa$ values (ie $\kappa \leq 0.01$ ) the energy straggling function is given by the Landau distribution (Leo, 1994).

The AlfaMC code has two options on what concerns the energy straggling. A first option uses only the Gaussian model. A second option uses the Gaussian/Vavilov/Landau models depending on the κ parameter. The latter option is more slower than the first one (three times slower on average). In most cases the difference introduced in the energy FWHM is small so the option to use only the Gaussian model is a reasonable one. This option is the program default. To choose between them the **istrag** flag is set via a call to the **alSetStrag** routine. This call can be made anytime during the program execution. Thus, the user has the ability to switch between straggling models during a run. The routines needed to generate the Vavilov and Landau distributions are not native to AlfaMC and are supplied in a separate library.

**2.2.1 The Gaussian distribution**

If the number of collisions of the alpha particle with the atomic electrons when traversing an absorber is large and each energy loss can be considered as independent then the Central Limit Theorem can be applied. A Gaussian distribution is then expected for the energy loss in the absorber with a certain standard deviation. Several forms have been proposed for this Gaussian straggling distribution. The first form was proposed by N. Bohr in 1913 (Bohr, 1913) and in his model the variance of the energy loss distribution is given as [Ramirez, 1969; Strittmatter, 1976; Chu, 1976]

$$\sigma_B^2 = 4\pi \left( \frac{e^2}{4\pi\varepsilon_0} \right)^2 z_\alpha^2 Z N \Delta x \qquad (5)$$

where e is the elementary charge, $\varepsilon_0$ the vacuum permittivity, N the number of atoms per unit volume, $\Delta x$ the slab thickness, Z the medium atomic number and $z_\alpha$ the alpha particle atomic number. Since $N = \rho N_A / A$ where ρ is the density, A the atomic weight and $N_A$ the Avogadro number, using the relation $r_e = (1/4\pi\varepsilon_0) e^2 / m_e c^2$ the variance can be written as

$$\sigma_B^2 = 4\pi r_e^2 (m_e c^2)^2 z_\alpha^2 \frac{Z}{A} \rho N_A \Delta x = 0.1569 z_\alpha^2 \frac{Z}{A} \rho \Delta x \, (\text{MeV}^2) \tag{6}$$

with $\Delta x$ in cm and $\rho$ in g/cm$^3$.

A quantum mechanical theory has been developed by Bethe and Livingston (Livingston 1937, Strittmatter, 1976; Tsoulfanidis, 1995) who found a value for the variance equal to

$$\sigma_{BL}^2 = 4\pi \left(\frac{e^2}{4\pi \varepsilon_0}\right)^2 z_\alpha^2 N \left( Z' + \sum_i \frac{8}{3}\left(\frac{I_i Z_i}{T_{max}}\right) \ln\left(\frac{T_{max}}{I_i}\right)\right) \Delta x \tag{7}$$

where Z' is the the effective atomic number (Strittmatter, 1976) and $I_i$ and $Z_i$ are the ionization potential and number of electrons of the i shell of the stopping atom. While the Bohr's result is independent of the particle energy, the Bethe-Linvingston's result has a small energy dependence, relevant for high energies (Tsoulfanidis, 1995).

More comprehensive theories have been developed for thick slabs such as the Payne (Payne, 1969) or Tschalar theories (Tschalar, 1970) but from the Monte Carlo point of view this theories have a more complex implementation and are not suitable for a fast Monte Carlo code. The GEANT3 and GEANT4 MC codes (GEANT, 1993; GEANT4, 2011) use an alternative formulation for the Gaussian variance which is based on the work of Seltzer and Berger (Schorr, 1974; Seltzer, 1964) which is suitable for a fast Monte Carlo. This formulation introduces a correction to the energy straggling variance depending on the particle relative velocity b and maximum energy transfer $T_{max}$

$$\sigma^2 = \frac{2\pi e^4 z_\alpha^2 Z \rho N_A \Delta x}{m_e c^2 \beta^2 A} T_{max}\left(1-\frac{\beta^2}{2}\right) = 0.1534 \frac{z_\alpha^2}{\beta^2}\frac{Z}{A} \rho \Delta x T_{max}\left(1-\frac{\beta^2}{2}\right) (\text{MeV}^2) \tag{8}$$

For alpha particles up to a few tens of MeV the correction introduced by this formula relative to $\sigma_B$ is negligible. In fact since the maximum energy transfer in a single collision is given by eq. 4 in the limit $m_e \ll m_\alpha$ and $\gamma \approx 1$ we get $T_{max} \approx 2m_e c^2 \beta^2 = 1.022 \beta^2$ MeV. Under these conditions the relative velocity is much less than the unit so that $(1-\beta^2/2) \approx 1$ and finally $\sigma^2 \approx \sigma_B^2$.

### 2.2.2 The Landau distribution

For very thin absorbers or gases the number of collisions can be too small for the Central Limit Theorem to hold. Large energy fluctuations are possible and the straggling function is no longer symmetrical. If the mean energy loss is approximated by the Bethe-Bloch constant term

$$\overline{E_{loss}} = \xi = \frac{2\pi z_\alpha^2 e^4 N_A Z \rho \Delta x}{m_e \beta^2 c^2 A} = 0.1534 \frac{z_\alpha^2 Z \rho \Delta x}{\beta^2 A} \quad \text{MeV} \tag{9}$$

the cut parameter will be given by $\kappa = \xi / T_{max}$. For $\kappa \leq 0.01$ the straggling distribution is successfully described by the Landau theory. For an energy loss $\varepsilon$ the Landau distribution $f(\varepsilon, \Delta x)$ may be written in terms of a universal $\phi(\lambda)$ function such as

$$f(\varepsilon, \Delta x) = \frac{1}{\xi} \phi(\lambda) \quad . \tag{10}$$

The Landau variable $\lambda$ is defined as (GEANT, 1993)

$$\lambda = \frac{\varepsilon - \overline{E_{loss}}}{\xi} - (1-C) - \beta^2 - \ln\left(\frac{\xi}{T_{max}}\right) \tag{11}$$

where C=0.577215... is the Euler's constant. In AlfaMC the $\phi(\lambda)$ is obtained from a modified version of the GEANT3 routine GLANDG (GEANT, 1993).

### 2.2.3 The Vavilov distribution

For intermediate thickness absorbers where $0.01 \leq \kappa < 10$ the energy loss distribution can be obtained from the Vavilov theory. This theory relates the energy loss distribution $f(\varepsilon,\Delta x)$ of a charged particle with an universal function $\phi_v(\lambda_v,\kappa,\beta^2)$ just in the same way as the Landau theory

$$f(\varepsilon,\Delta x)=\frac{1}{\xi}\phi(\lambda_v,\kappa,\beta^2) \; , \tag{12}$$

where the Vavilov variable $\lambda_v$ is defined as (GEANT, 1993)

$$\lambda_v=\kappa\left(\frac{\varepsilon-\overline{E_{loss}}}{\xi}-(1-C)-\beta^2\right) \; . \tag{13}$$

The Vavilov $\lambda_v$ is related to $\lambda$ by the relation (GEANT, 1993) $\lambda=\lambda_v/\kappa-\ln\kappa$ and the relation between $\lambda$ and the energy loss is

$$\lambda_v/\kappa=\lambda+\ln\kappa=\left(\frac{\varepsilon-\overline{E_{loss}}}{\xi}-(1-C)-\beta^2\right) \; . \tag{14}$$

AlfaMC uses the GVAVIV routine from GEANT3. This routine samples the $\lambda$ Landau instead of the $\lambda_v$ variable.

**2.3 Alpha particle multiple scattering**

An alpha particle traversing a medium is deflected through many small-angles, mostly due to Coulomb scattering from nuclei. The distribution of the scattering angle after traversing a small layer is roughly Gaussian for small angle values. At larger angles (greater than a few standard deviations) the distribution has a Rutherford scattering-like behavior, with larger tails than those of a Gaussian distribution. The Molière's (Bethe 1953) and Fermi's (Rossi, 1941) theories have been widely used to describe the multiple scattering of heavy charged particles. The theory of Fermi results in the Gaussian approximation for small angles, has an intuitive physical meaning and is easy to implement in the MC code. For these reasons was the adopted as the multiple scattering model in AlfaMC.

For a particle traversing a thin slab of matter of thickness s with incidence direction along the z axis, one can define the deflection angles $\theta_x$ and $\theta_y$ (figure 1), measured relatively to the incidence direction in the xz and yz planes. According to the Fermi's theory (Beringer, 2012; Wong, 1990) the $\theta_x$ and $\theta_y$ deflection angles have independent Gaussian distributions given by

$$\frac{dN}{d\theta_i} = \frac{1}{\sqrt{2\pi}\,\theta_0} \exp\left(-\frac{\theta_i^2}{2\theta_0^2}\right) \qquad (15)$$

with i=x,y.

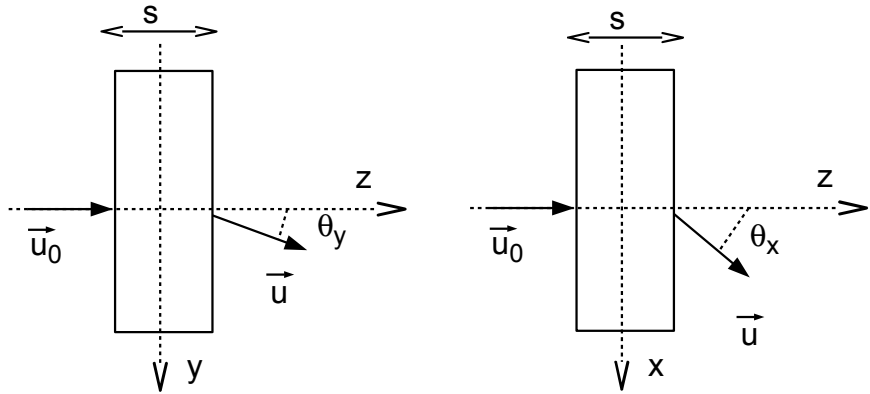

Figure 1. Definition of the $\theta_y$ and $\theta_x$ deflection angles.

The standard deviation of the distribution $\theta_0$ can be approximated by (Wong, 1990; Beringer, 2012)

$$\theta_0 = \frac{13.6\,\text{MeV}}{pc\,\beta} Z_\alpha \sqrt{\frac{s}{X_0}} \qquad (16)$$

where $X_0$ is the material radiation length. Lynch and Dahl (Lynch, 1991) have looked into to the problem of parameterizing $\theta_0$ and propose several approximations. The adopted by AlfaMC was

$$\theta_0 = \frac{4}{5} \cdot \frac{13.6 \, \text{MeV}}{pc\beta} Z_\alpha \sqrt{\frac{s}{X_0}} \left[ 1 + 0.038 \ln\left( \frac{s Z_\alpha^2}{X_0 \beta^2} \right) \right] \qquad (17)$$

where the 4/5 factor was empirically determined, for better comparison with the SRIM results (see section on comparison with SRIM).

The material radiation length $X_0$ has been calculated and tabulated by Y.S. Tsai (Tsai, 1974) and can be given by the formula

$$\frac{1}{X_0} = 4\alpha r_e^2 \frac{N_A}{A} \left\{ Z^2 [L_{rad} - f(Z)] + Z L'_{rad} \right\} \quad \text{g cm}^{-2} \qquad (18)$$

where $\alpha$ is the fine structure constant, $r_e$ the electron classical radius, $N_A$ the Avogadro number, A the mass number, and Z the atomic number. For A=1 g mol$^{-1}$, $4\alpha r_e^2 N_A / A = (716.408 \, \text{g cm}^{-2})^{-1}$ when expressing $X_0$ in g cm$^{-2}$. The function f(z) can be approximated for elements up to uranium by (Beringer, 2012)

$$f(Z) = a^2 \left[ (1+a^2)^{-1} + 0.20206 - 0.0369 a^2 + 0.0083 a^4 - 0.002 a^6 \right] \qquad (19)$$

where $a = \alpha Z$ (Davies, 1954). The functions $L_{rad}$ and $L'_{rad}$ are given in table 1.

Table 1. Tsai's $L_{rad}$ and $L'_{rad}$ values used in the calculation of the radiation length in an element.

| Element | Z | $L_{rad}$ | $L'_{rad}$ |
|---|---|---|---|
| H | 1 | 5.31 | 6.144 |
| He | 2 | 4.79 | 5.621 |
| Li | 3 | 4.74 | 5.805 |
| Be | 4 | 4.71 | 5.924 |
| Others | >4 | ln(184.15 Z$^{-1/3}$) | ln(1194 Z$^{-2/3}$) |

Equation 16 is applicable to one path of length t. Dividing the path into steps (i.e. $s=s_1+s_2$) will introduce a bias since $\theta_0(s_1+s_2) \neq \sqrt{\theta_0^2(s_1)+\theta_0^2(s_2)}$. AlfaMC neglects this effect, introducing a systematic bias in the lateral straggling.

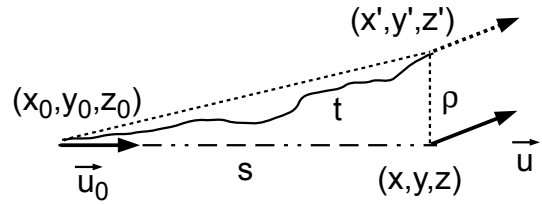

Figure 2. Relation between the particle step s and the actual path length t.

In Monte Carlo transport of an alpha particle, due to the multiple scattering the actual path length t made by the alpha particle is bigger than the step s (figure 2). From the Fermi-Eyges theory we have the relation (Nelson, 1985)

$$t = s + \frac{1}{2}\int_0^t \theta_0^2(t')\,dt' \qquad (20)$$

where $\theta_0(t)$ is given by eq. 16. Assuming a constant momentum p during the step and approximating t by s in the integral one gets

$$t \approx s + \frac{K}{4}s^2 \qquad (21)$$

where $K=[(13.6\,\text{MeV})Z_\alpha/pc\beta]^2/X_0$ and $X_0$ in cm. For 1 μm steps or smaller the correction is less than 0.1% for low Z materials and energies above 1 MeV. The correction becomes more important as the alpha particle decreases its energy and $\theta_0(t)$ increases. The correction might also be important for high-Z material where $\theta_0(t)$ also has a larger value due to the increase in the Coulomb cross-section. The effect enters the used formulas through the decrease of $X_0$ with Z which can be observed in the approximate equation (Beringer, 2012)

$$X_0 = \frac{716.4 \, \text{gcm}^{-2} A}{Z(Z+1)\ln(287/\sqrt{Z})} \tag{22}$$

In AlfaMC the K factor is computed as $K = \theta_0^2/s$ and $\theta_0$ is taken from eq. 17.

Apart from the path length correction, there is also a transverse correction. In fact the particle position after a step s is computed as $(x,y,z) = (x_0, y_0, z_0) + s\vec{u}_0$ where $\vec{u}_0 = (u_0, v_0, w_0)$ is the particle's direction vector entering the slab. Due to multiple scattering the particle position should be $(x', y', z')$, a distance $\rho$ from $(x,y,z)$. Since t is bigger than the triangle hypotenuse we can say that $\rho^2 < t^2 - s^2 \approx (s + K/4s^2)^2 - s^2 = K/2 s^3 + (K/4)^2 s^4$. As far as s stays small and $t \approx s$ the lateral displacement $\rho$ is very small and can be neglected.

In the laboratory reference frame upon entering the slab the particle's direction vector is $\vec{u}_0 = (u_0, v_0, w_0)$. The $\theta_x$ and $\theta_y$ are generated in the reference frame where the particle initial direction is assumed to be the z axis. In this frame the particle's initial direction is given by the the vector $\vec{q}_0 = \vec{e}'_z$ and the new direction after traversing the slab is $\vec{v} = v_x \vec{e}'_x + v_y \vec{e}'_y + v_z \vec{e}'_z$ where

$$\begin{aligned} v_x &= \sin\theta_x \\ v_y &= \sin\theta_y \\ v_z &= \sqrt{1 - v_x^2 - v_y^2} \end{aligned} \tag{23}$$

The model is valid for small $\theta_x$ and $\theta_y$ but since they are independently sampled there is a (small) probability that $v_x^2 + v_y^2 > 1$. To ensure this is not the case the module of vector $\vec{v}$ is computed and the direction cosines obtained as

$$\cos\alpha = \vec{v}_x/|\vec{v}|$$
$$\cos\beta = \vec{v}_y/|\vec{v}| \qquad (24)$$
$$\cos\gamma = \vec{v}_z/|\vec{v}|$$

and the direction vector is $\vec{v} = \cos\alpha\, \vec{e}'_x + \cos\beta\, \vec{e}'_y + \cos\gamma\, \vec{e}'_z$.

We notice the choice of the x and y axis in this reference frame is arbitrary as long as the x,y,z axis are orthogonal to each other. A possible choice is to define the x axis in the particle reference frame as $\vec{e}'_x = \vec{e}'_z \times \vec{e}_z$ and the y axis as $\vec{e}'_y = \vec{e}'_z \times \vec{e}'_x$.

Using this reference frame the rotation of the direction vector to the laboratory frame is given by

$$u = (v_0/D)\cdot\cos\alpha + (u_0 * w_0)/D \cdot \cos\beta + u_0 \cdot \cos\gamma$$
$$v = (-u_0/D)\cdot\cos\alpha + (v_0 * w_0)/D \cdot \cos\beta + v_0 \cdot \cos\gamma \qquad (25)$$
$$w = 0\cdot\cos\alpha - D\cdot\cos\beta + w_0 \cdot \cos\gamma$$

where $D = \sqrt{(u_0^2 + v_0^2)}$.

**2.4 Alpha-particle effective charge**

While slowing down the alpha-particles capture electrons from the medium reducing the effective electrical charge.
Several author have studied the effect putting forward a number of parameterizations. In a simple model the effect can be parametrized in terms of the reduced particle velocity

$$Z_{eff} = Z[1 - \exp(0.92 u/(v_0 Z^{2/3}))]$$

where $v_0$ is the Bohr electron velocity $v_0 = \dfrac{e^2}{4\pi\varepsilon_0 \hbar} = \alpha c = \dfrac{c}{137}$ and $u = \beta c$, so that (Hatano 2011)

$$Z_{eff} = Z[1 - \exp(125\beta/Z^{2/3})]$$

The effect is already included in the NIST stopping-power, but not in the energy straggling and multiple-scattering since the adopted models tend to give smaller values than the ones foreseen by more realistic models.

**3. The program flow**

The AlfaMC code is written in Fortran language and the main code is found in the library AlfaMCLIB.f . It uses the ULYSSES (Ulysses, 2012) package for the particle tracking in the geometry and the ULHISTOS package for scoring and histogramming (Ulhistos, 2012). ULYSSES is a package, designed to make particle tracking in complex volumes and score the results. The histogramming routines are grouped in a library called ULHISTOS. The ULYSSES package is written in FORTRAN and some knowledge of the language is necessary for an efficient usage of the program. There are several volumes available in ULYSSES that can be used to build complex bodies. Examples of available volumes are the rectangular box, the cylinder, the sphere, the cone, etc. Volumes can be rotated in space. The geometry system used by ULYSSES allows for the construction of rather complex structures by adding (or subtracting) volumes. The volume organization is made using mother-daughter logic. Any volume may have daughter volumes inside. All the volumes have a mother volume except the universe volume which contains all volumes. A volume may have more than one mother, that is, it may be shared by more than one volume. The scoring of results is handled by the ULHISTOS library. One or two-dimensional histograms are possible. Apart from booking, filling and outputting histograms, the ULHISTOS library contains several routines to perform operations on the histograms and to extract statistical results.

The code is steered by a main routine which controls the program flow. Examples of the main.f program are supplied in the examples folders but the code is open to changes by the user. A routine containing the geometry description must be supplied by the user. This routine (ulgeom.f) uses the ULYSSES package routines to build the geometrical setup. A radiation source routine containing the code for the generation of the alpha particle initial parameters (e.g. position, direction and energy) must also be supplied by the user. Examples of such routines can be found in the Examples folder.

The flowchart of the main routine is presented in the figure 3. The program starts by initializing general run parameters in the ULYSSES and ULHISTOS databases. The run parameters (number of events to be generated and number of materials used) and cuts (energy and step size) are set. The necessary histograms to score the results are then booked. The data files containing the stopping power tables and other data characterizing the material media are also read. Then the routine ulgeom.f containing the geometrical description of the setup is called. After that the program enter the main event loop where each alpha particle is generated and tracked through the geometry. The generation of the alpha particle initial properties (position, direction and kinetic energy) is made in routine ulsource.f that must be supplied by the user. The program then finds the volume inside which the particle is generated. Then a step size is computed according to the percentage of energy loss allowed for each step and set by variable dEstep. This variable must have a value between 0 and 1. The next step size s is computed according to $s = dxdE \times (E \cdot dEstep)$ where dxdE is the reciprocal of the stopping power. If the computed value is lower or higher than the step size cuts defined by the user the step size reverts to the limit. To ensure proper randomness in successive events a Gaussian sampling of the step size is further done with a standard deviation fixed in 20% of the precomputed step size. The particle is then tracked through the geometry. Starting at the current $\vec{x}_0$ position, a new position along direction $\vec{u}$ is found as $\vec{x} = \vec{x}_0 + s \cdot \vec{u}$. If a volume boundary is crossed the step size is shortened so that the particle is left at the boundary but on the new volume side.

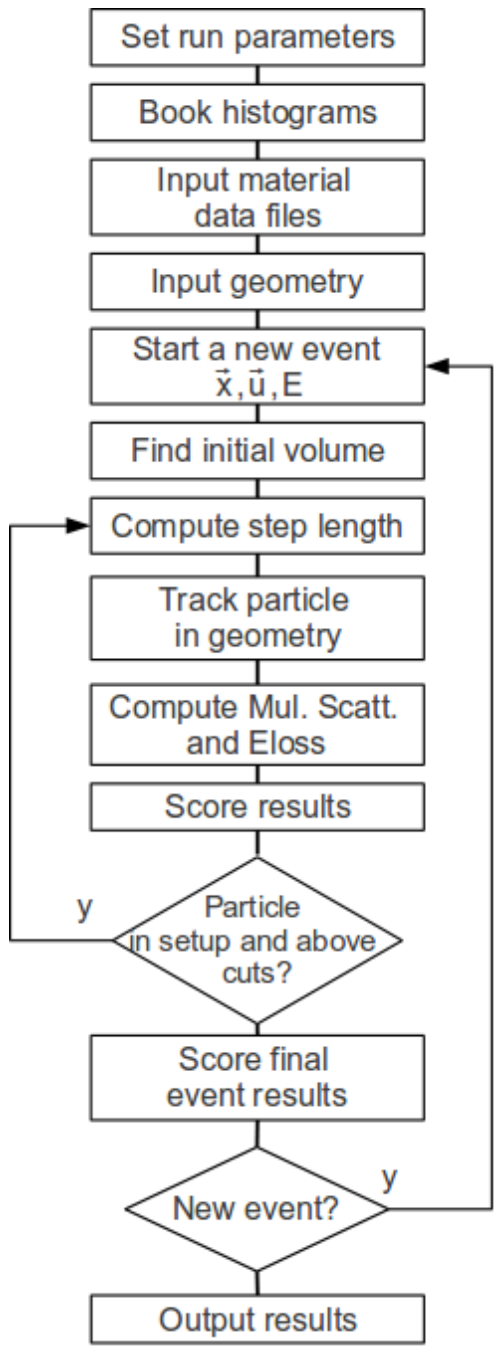

Figure 3. The program flowchart.

After determining the particle's new position the a new particle direction $\vec{u}'$ due to multiple scattering is computed. This direction will be assumed to be the particle's fight direction after the step is completed. A correction to the step length due to multiple scattering is made and the energy loss in the step computed. Results such particles fluencies or deposited energy in the tracking volume can be scored. When crossing volume boundaries, the deposited energy in the step must be assigned to the previous volume since the particle is left on the boundary line between volumes. After the step the program checks if the particle still has the energy above the energy cut and did not exit the setup. If both requirements are met the program will advance the particle one more step. Otherwise, final event results (like the total deposited energy in a volume) are scored, and the program will generate a new particle. When the total number of requested particles are generated the program outputs the final results.

**4. The materials database**

The AlfaMC package provides the program AlfaMaterial.f to compute stopping-power data. The program uses the data from the NIST/ASTAR (Berger, 2011b) database for the computation of the stopping-power. The ASTAR program calculates stopping-power and range tables for helium ions in 74 materials, according to methods described in ICRU Report 49 (ICRU, 1993) and briefly explained in the NIST/ASTAR website page (NIST, 2012a). The list includes 26 elements and 48 compounds and mixtures. The available energy ranges between 0.001 and 1000 MeV. The package provides a folder (matdb) containing the data files with the electronic and nuclear stopping-powers and ranges. For compounds and mixtures an additional file contains the material composition. The AlfaMaterial program can compute the stopping-power of user-defined materials using the Bragg rule (NIST, 2012b, ICRU, 1993):

$$\frac{1}{\rho}\frac{dE}{dx} = \sum \frac{w_i}{\rho_i}\left(\frac{dE}{dx}\right)_i . \qquad (26)$$

where $w_i$ and $\rho_i$ are the fraction by weight and density of the ith element.

Besides stopping-power, AlfaMaterial computes other useful parameters for the tracking of alpha particles, namely the radiation length $X_0$, the mean excitation energy, the effective atomic number and effective mass number. The radiation length $X_0$ is computed according to equation 18 (Beringer, 2012, Tsai, 1974).

The radiation length in a compound or mixture may be approximated by

$$\frac{1}{X_0} = \sum \frac{w_j}{X_{0j}} \tag{27}$$

where $w_j$ and $X_{0j}$ are the fraction by weight and radiation length of the jth element.

The mean excitation energy value I for elements is obtained from the table available at NIST (NIST, 2012b). For a compound or mixture the mean excitation energy value can be approximated by (Attix, 2008, Geary, 1976)

$$\ln(I) = \frac{\sum f_i (Z_i/A_i) \ln(I_i)}{\sum f_i (Z_i/A_i)} \tag{28}$$

where $f_i$ is the proportion by weight of element i. In the case of a compound

$$f_i = \frac{N_i A_i}{\sum N_i A_i} \tag{29}$$

where $N_i$ is the number of atoms of element i in the compound. The I value obtained with this formula will be within 5% of the value quoted by NIST for low effective atomic number compounds ($Z_{eff} < 8$) while deviations up to 15% can be obtained for higher $Z_{eff}$. The impact of this discrepancies on the AlfaMC results in the present 2.0 version is zero since this parameter is not used in the simulation

The atomic weights of elements are obtained from (Atomic weights, 2009; Wieser, 2011).

## 5. Geometry and histogramming

A strong feature in AlfaMC is the ability to simulated complex geometrical setups and to score results using the ULYSSES and ULHISTOS packages (Ulysses, 2012 Ulhistos, 2012). To use the ULYSSES package, the user must supply a geometry and a source routine, along with the main program. A set of predefined volumes (boxes, ellipsoids, tubes, etc.) can be combined to build complex geometrical bodies, where the particles will be tracked. In the source routine, also supplied by the user, the starting position, direction and energy of the particle is generated.

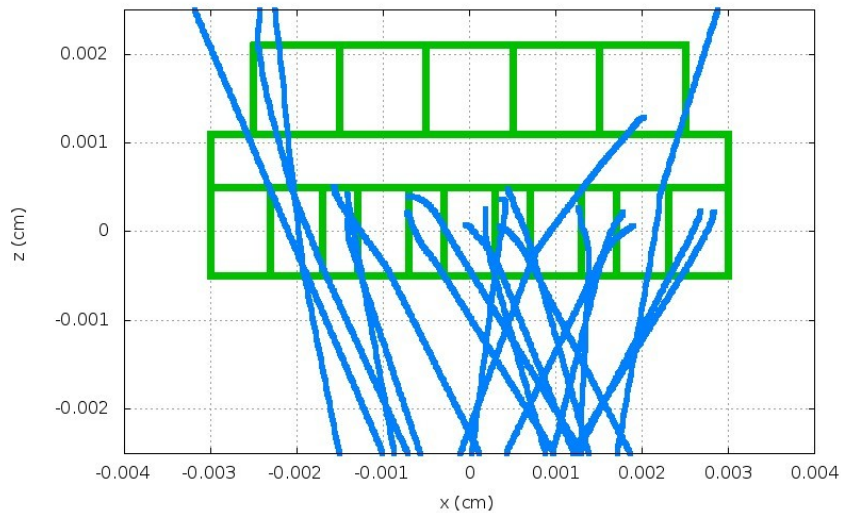

Figure 4.  5 MeV alpha particles are emitted from an extended isotropic source. A 10 μm-thick gold collimator is placed between the source and a layer of cells placed over a 6 μm-thick mylar foil.

An example is displayed in the figure 4. An extended source in air, emits 5 MeV alpha particles isotropically. For the sake of clarity, alpha particles emitted at large angles have been eliminated from the figure. A 10 μm-thick gold collimator is placed between the source and a layer of cells. The collimator's holes are 3 μm in diameter. A layer of cells is placed over a 6 μm-thick mylar foil. The cells are 10 μm-thick and the material is water. A 5 MeV alpha particle has a projected range in gold of 8.3 µm so the collimator is thick enough to stop the alpha particles crossing it. Only alpha particles passing through the holes have a chance to hit the cells sitting on the top of the mylar foil.

## 6. Comparison with SRIM

The SRIM program (SRIM, 2008) contains updated values of the stopping powers for most ion nuclei including helium ions. Using a simple slab geometry several physical quantities can be compared when computed by AlphaMC and SRIM. The comparison will be made for 4 different media: a gas (air) 5 mm thick, a low Z medium with absorption properties close to human tissue (mylar) 3 µm thick, a medium Z absorbent material (aluminum) 2 µm thick and a high Z material (gold) 1 µm thick. Thicknesses where chosen in such a way that, percentage energy loss is of same order of magnitude for all 4 targets. For each data point a run of 10000 events was made.

### 6.1 Transmitted energy

The transmitted energy of alpha particles (figure 5) traversing different thicknesses of the four chosen media is computed by AlfaMC and SRIM program. The thicknesses where chosen in such a way that, depending on the beam energy, the layer could represent alpha particle behavior from a thick to a thin absorber. The energy ranges from 0.1 to 12 MeV. The average value and standard deviation (energy straggling) were computed using a $3\sigma$ interval relatively to the full distribution. This procedure minimizes the bias due to non-Gaussian tails.

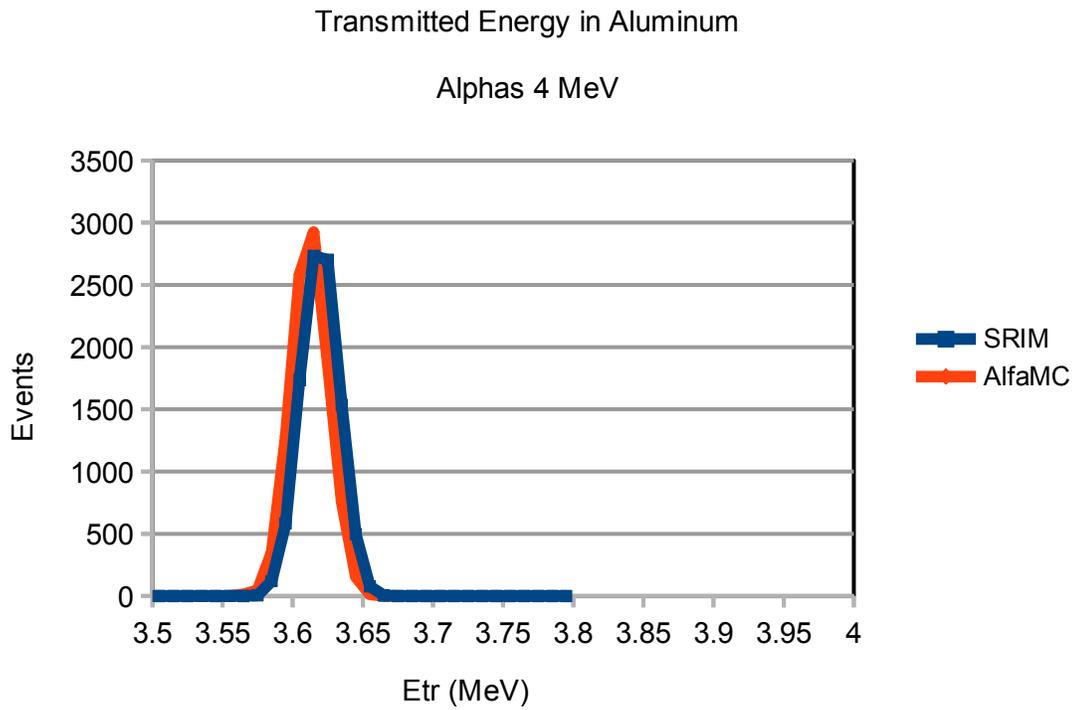

Figure 5. Transmitted energy in aluminum for 4 MeV alpha particles simulated by SRIM and AlphaMC.

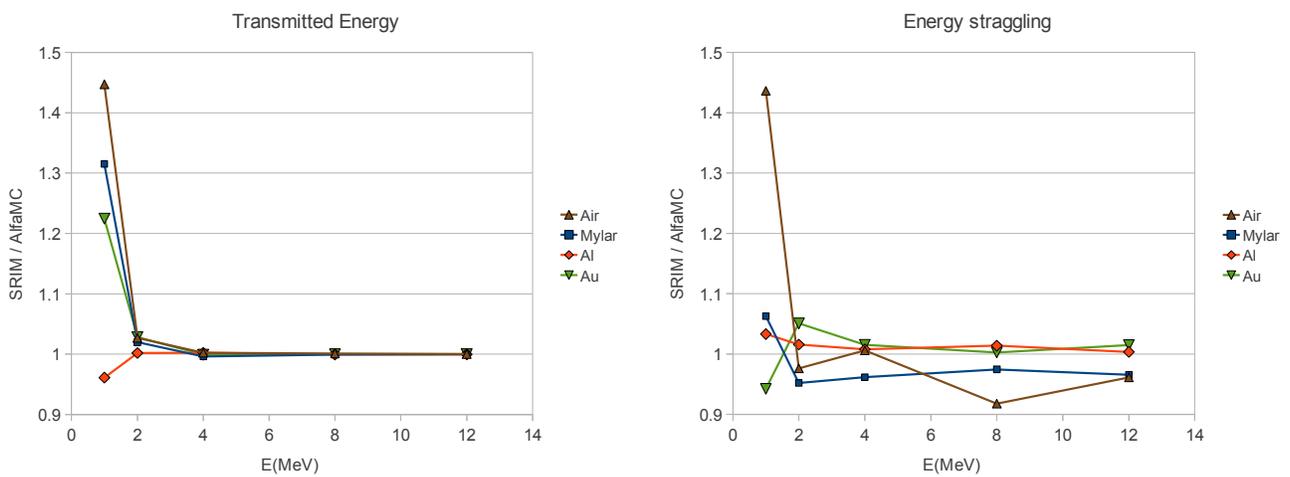

Figure 6. Ratio of alpha particle transmitted energy and energy straggling for 4 materials.

Except for the cases of low energy beam where the transmitted energy is very low the agreement between SRIM and AlfaMC is within 1 to 2% that is the same uncertainty as the computation statistical error (figure 6). If instead of the transmitted energy, we analyze the energy lost (deposited) by the alpha particle in the layer then the agreement in the low energy is better than 10%. For the energy straggling (measured by the energy standard deviation) the general agreement between SRIM and AlfaMC is within 6%.

**6.2 Lateral dispersion**

The value of the lateral dispersion on the coordinates perpendicular to the beam axis (standard deviation on x or y) is a measure of the multiple scattering lateral straggling. Four different parameterizations of the polar scattering angle have been considered (Lynch, 1991). Three of them are variations of the Highland formula (Highland, 1975; Highland, 1979)

$$\theta_{01} = \frac{13.6\,\text{MeV}}{pc\,\beta} Z_\alpha \sqrt{\frac{s}{X_0}} \quad , \tag{30}$$

$$\theta_{02} = \frac{13.6\,\text{MeV}}{pc\,\beta} Z_\alpha \sqrt{\frac{s}{X_0}} \left[1 + 0.038 \ln\left(\frac{s}{X_0}\right)\right] \quad , \tag{31}$$

$$\theta_{03} = \frac{13.6\,\text{MeV}}{pc\,\beta} Z_\alpha \sqrt{\frac{s}{X_0}} \left[1 + 0.038 \ln\left(\frac{s Z_\alpha^2}{X_0 \beta^2}\right)\right] \quad , \tag{32}$$

and a four formula proposed by Lynch and Dahl (Lynch, 1991) and modified in Geant3 manual (GEANT, 1993)

$$\theta_{04}^2 = \frac{\chi_c^2}{1+F^2}\left[\frac{1+\nu}{\nu}\ln(1+\nu) - 1\right] \tag{33}$$

where F=0.98,

$$\nu = 0.5\Omega/(1-F) \;,$$

$$\Omega = \frac{\chi_c^2}{1.167\chi_\alpha^2} \;,$$

$$\chi_c^2 = 0.157[Z(Z+1)X/A][z_\alpha/(pc\beta)]^2 \;,$$

$$\chi_\alpha^2 = 2.007 \times 10^{-5} Z^{2/3}[1+3.34(Zz_\alpha\alpha/\beta)^2]/(pc)^2 \;\; \text{and}$$

$$\alpha = 1/137 \;.$$

For air, mylar aluminum and gold the value of the lateral dispersion has been computed by AlfaMC and compared to the value obtain in the SRIM simulation in the 1 to 12 MeV range. The obtained results are presented in figure 7.

We conclude that none of the models give a satisfactory description of the multiple scattering lateral straggling for the entire studied range. The values of $\theta_{02}$ and $\theta_{04}$ ratios have a strong energy dependence which is undesirable. On the other hand the values of $\theta_{01}$ and $\theta_{03}$ have a more smooth energy dependence, but giving too high values for the lateral straggling. For $\theta_{03}$ this higher value can be compensated introducing an empirical factor of 4/5 in the formula.

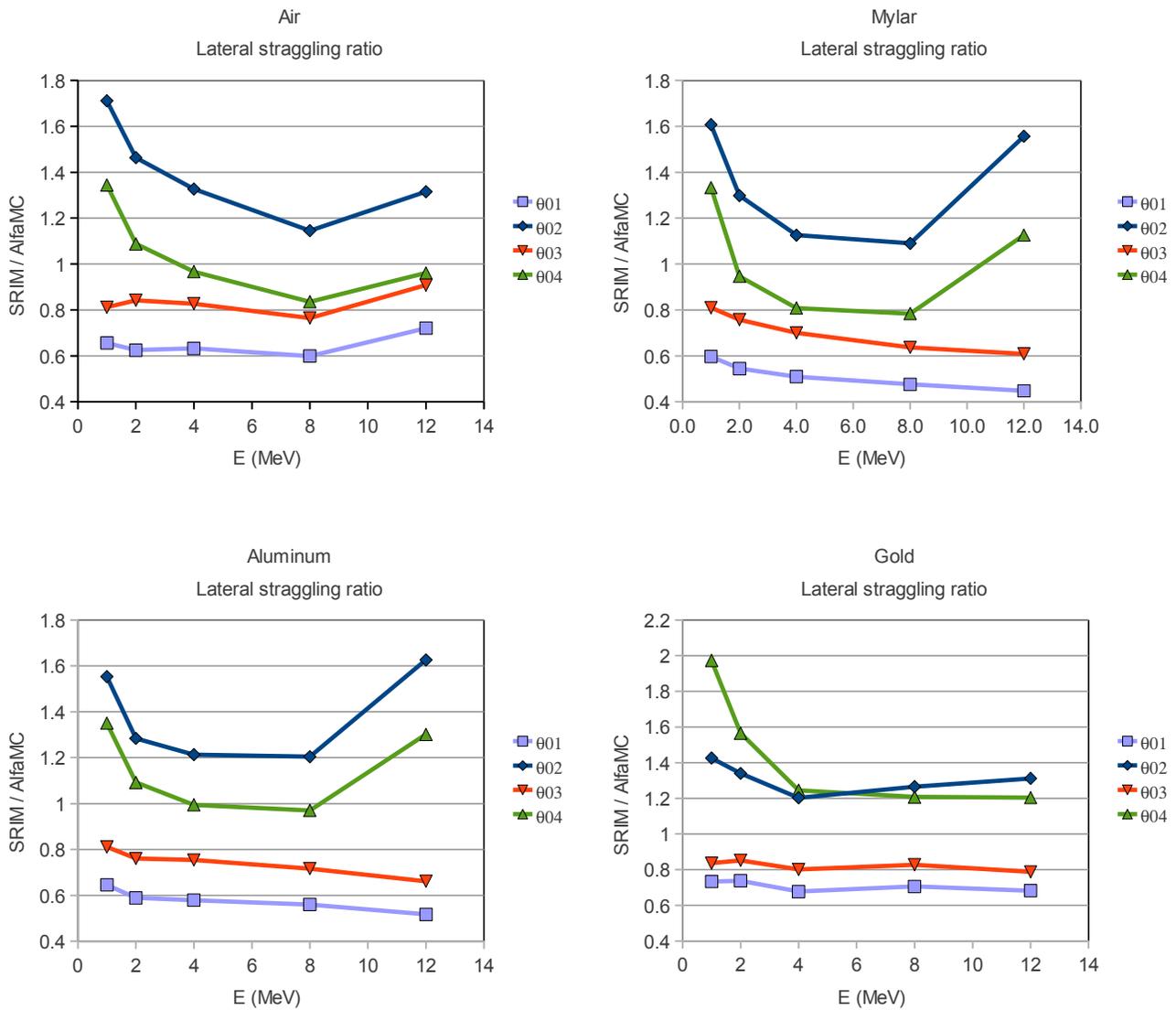

Figure 7. The SRIM / AlfaMC ratio for the lateral straggling in air (5 mm) , mylar (3 µm), aluminum (2 µm) and gold (1 µm) in the 1 to 12 MeV range.

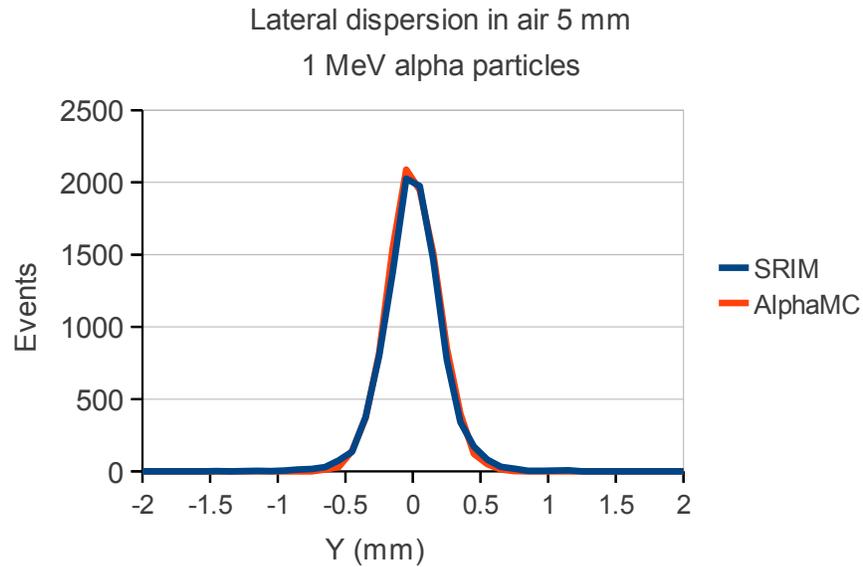

Figure 8. Lateral dispersion of 1 MeV alpha particles in 5 mm of air.

For a layer of 5 mm of air the lateral dispersion of alpha particles is presented in the figure 8. A good agreement is seen between the results of the two simulation programs.

**6.3 Range of alpha particles**

There is more than one quantity that can be defined under the name of *range* of a charged particle. Unfortunately authors are not always clear about to which type of range they are referring to. F. Attix (Attix, 2008) defines *range* as "The range R of a charged particle of a given type and initial energy in a given medium is the expectation value of the pathlength p that it follows until it comes lo rest (discounting thermal motion)". This is clearly a difficult quantity to measure, although being easily obtained from the computational point of view.

A closely related quantity is the called *CSDA range* (Attix, 2008; Johns 1983) representing the pathlength of the particle in the CSDA approximation. This quantity can be computed from the stopping power by

$$R_{CSDA} = \int_0^{T_0} \left(\frac{dE}{dx}\right)^{-1} dE \tag{34}$$

where $T_0$ is the particle initial kinetic energy. The above defined range R and the *CSDA range* $R_{CSDA}$ have almost identical values for heavy charged particles. The small difference comes from the occurrence of discrete and discontinuous energy losses along the path. The *CSDA range* is a quantity that can be computed with AlfaMC.

Another defined quantity is the *projected range*. As defined by F. Attix (attix, 2008) "the projected range <t> of a charged particle of a given type and initial energy in a given medium is the expectation value of the farthest depth of penetration of the particle in its initial direction". This is a quantity that can be experimentally determined in a transmission experiment. We can think of a pencil beam of particles perpendicularly traversing a variable thickness slab and measuring the number of particles that come out of the slab as we increase the slab thickness. For heavy charged particles and excluding the cases of nuclear interaction, almost all of them will transverse the absorbing material until a certain thickness $t_1$ is reached. After that thickness a rapidly decrease on the detected particles is observed. The thickness $t_{max}$ for which no charged particle come out of the slab is called the *maximum penetration depth*. The projected range can be obtained as

$$\langle t \rangle = \int_{t_1}^{t_{max}} t \frac{dN(t)}{dt} dt \ / \int_{t_1}^{t_{max}} \frac{dN(t)}{dt} dt \tag{35}$$

where $dN(t)/dt$ is the number of particles stopping at depths between t and t+dt. For practical reasons the *projected range* can be obtained from the experimental curve as slab thickness for which the beam intensity reduces to 50% of the plateau value . This range value is also known as mean range $R_{50}$ (Knoll, 2010; Johns, 1983). Yet another quantity called the *extrapolated range* $R_e$ can be defined. This is obtained by extrapolating at $R_{50}$ the nearly linear portion of the end of the transmission curve to zero (Knoll, 2010). From their definition it is clear that the relation $\langle t \rangle \simeq R_{50} < R_e < R$ must exist.

All this quantities can be obtained by simulation, in particular by AlfaMC. The *CSDA range* is obtained as the average over the number of simulated alpha particles of the sum of all steps given during the transport of each alpha particle. The program uses the multiple scattering corrected step to obtain the value of the *CSDA range*. The *projected range* is obtained by AlfaMC as the average depth of penetration of the alpha particles up to their full stop.

The *projected range* was computed by AlfaMC and SRIM for air, mylar, aluminum and gold for a number of energies in the range 0.1 to 12 MeV using $10^4$ events in each case. Figure 9 shows the ratio of the *projected range* as computed by SRIM and AlfaMC as a function of the particle kinetic energy. The ratio of the *projected range* given by SRIM and NIST in their original tables is also presented.

We start noticing the differences between the projected ranges given by SRIM and the NIST/ASTAR tables for low energies, that can be as much as 9%. On what concerns AlfaMC, from these results we conclude that except for low alpha energies (< 0.5 MeV) the computed range ratio SRIM/AlfaMC agrees with the SRIM/NIST range ratio within 2 to 3% at energies up to 2 MeV and is better than that for higher energies. The fact AlfaMC gives a higher error at low energies is related to the fact the computation is based on a set of given values in a table. At low energy the number of independent table values entering the computation is limited and the uncertainties add up.

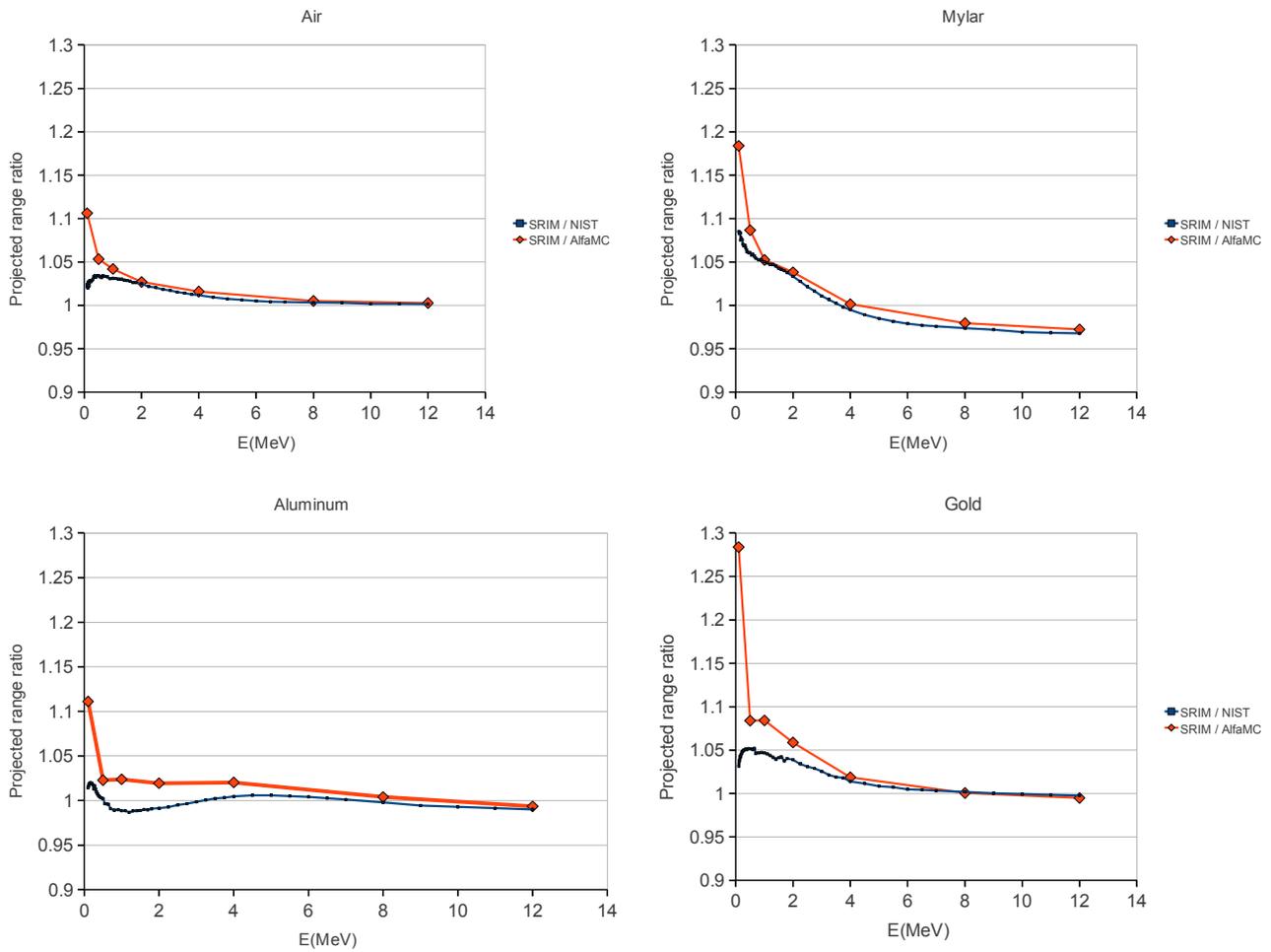

Figure 9. Projected range ratios between SRIM and AlfaMC for air, mylar aluminum and gold.

The *project range* curves in air for 1 MeV alpha particles for SRIM and AlfaMC is presented in the figure 10. There is, in this case, a 4% difference between the average value given by AlfaMC and SRIM, which can be assigned to the underlying difference between the SRIM and ASTAR/NIST values. A larger difference exist between the standard deviation obtained by each of the programs being the SRIM value 60% larger in this case. As it can be seen from figure 11, except for high energy values, the range straggling in SRIM has larger values than in AlfaMC. This can in part be due to the non-gaussian high angle multiple scattering which is not present in AlfaMC.

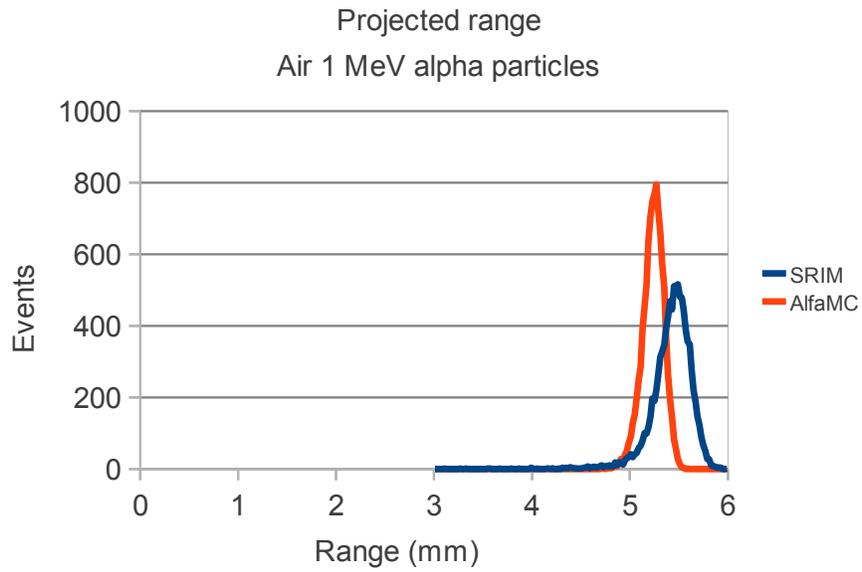

Figure 10. Projected range in air for 1 MeV alpha particles for SRIM and AlfaMC.

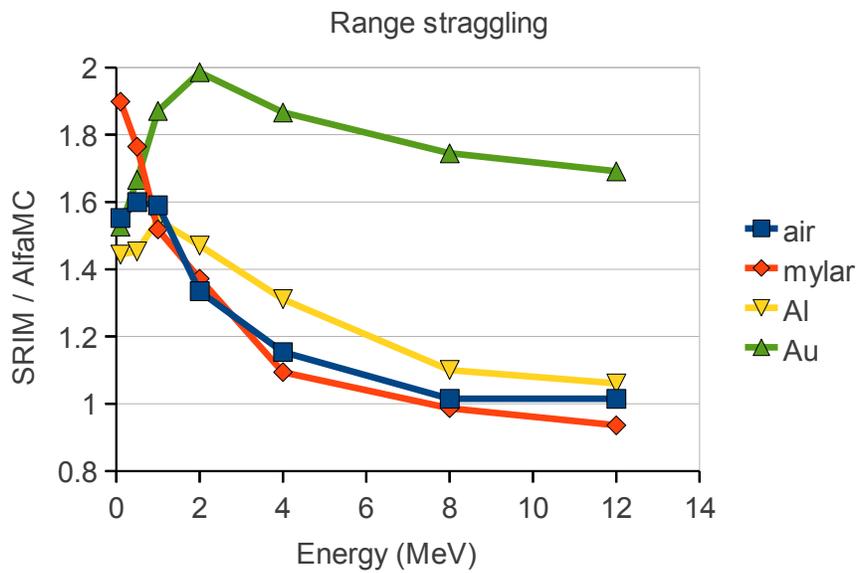

Figure 11. Range straggling parameter ratios between SRIM and AlfaMC for air, mylar, aluminum and gold.

This difference in the multiple scattering models is particularly important at the end of the alpha particle path where their energy is low. As expected the lateral dispersion at the end of the path has larger values for SRIM. This can be confirmed in figure 12 where the ratios between the standard deviations of the lateral dispersion curves are presented for air, mylar, aluminum and gold.

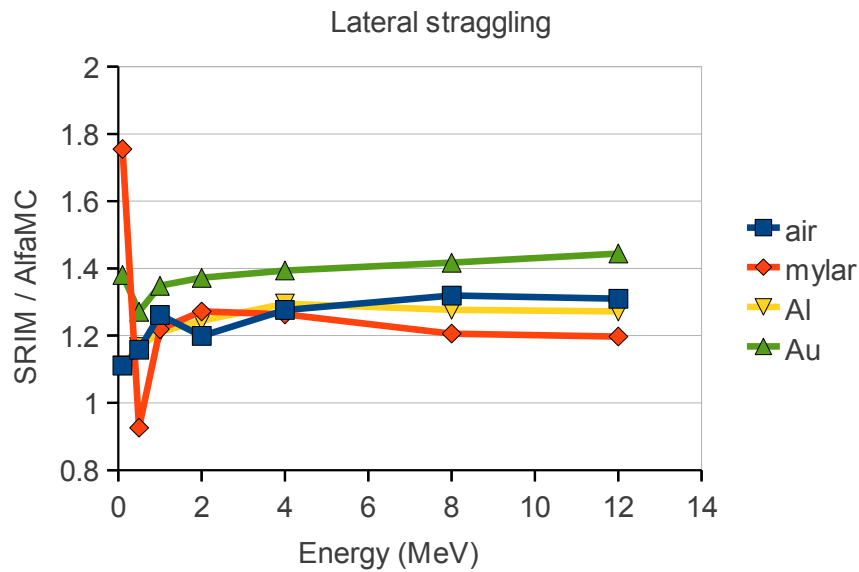

Figure 12. Lateral straggling parameter ratios between SRIM and AlfaMC for air, mylar, aluminum and gold.

For 1 MeV alpha particles in air the dispersion curve for SRIM and AlfaMC is presented in figure 13. Figure 14 shows the AlfaMC simulation of the transport of 0.5 MeV alpha particles in air.

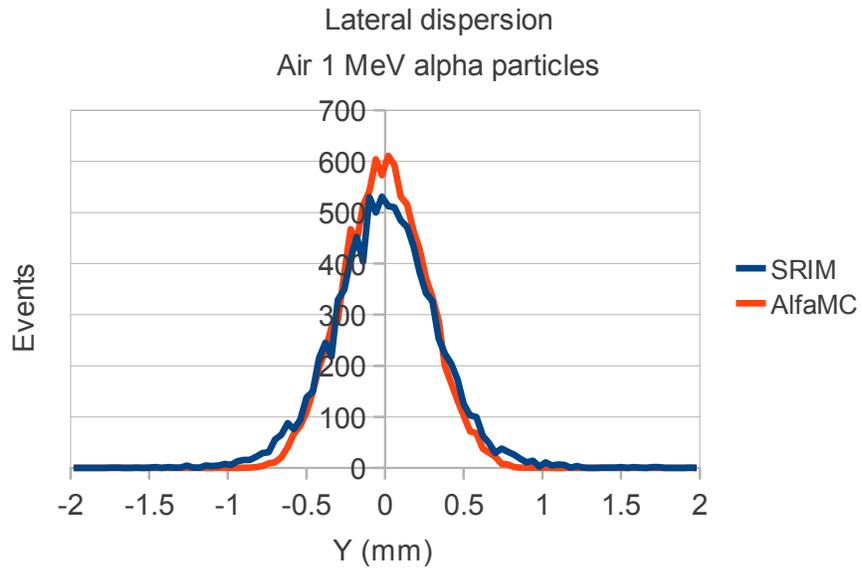

Figure 13. Lateral dispersion in air at the end of the path of 1 MeV alpha particles for SRIM and AlfaMC.

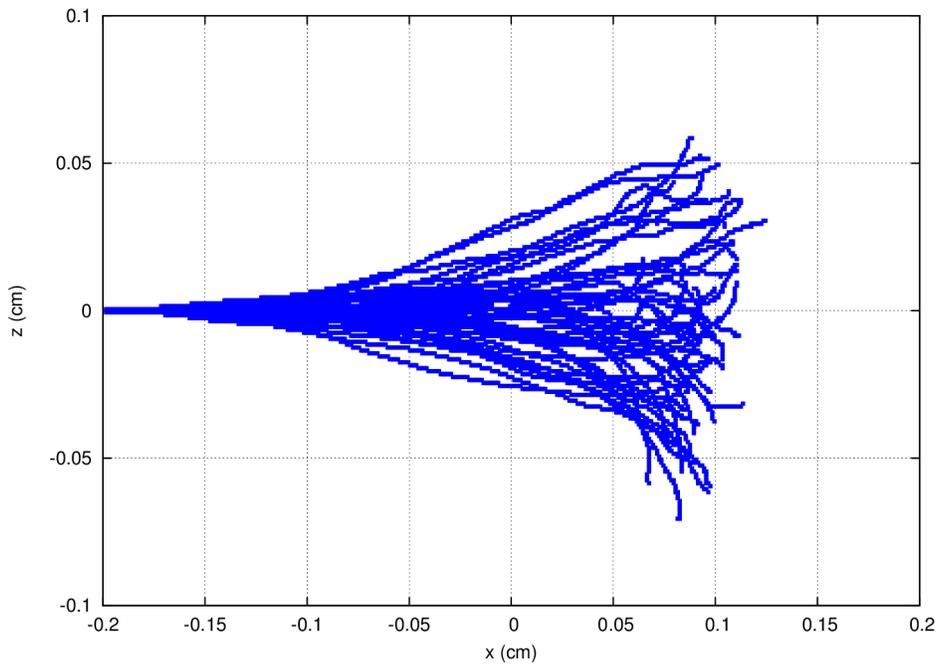

Figure 14. Simulation of a pencil beam of 0.5 MeV alpha particles in air by AlfaMC.

# 7. Conclusions

AlfaMC is a fast MC for the transport of alpha particles in complex geometries. The program uses the ASTAR / NIST precomputed stopping power tables to get the alpha particles energy loss. The default energy straggling is assumed to be Gaussian, but a more comprehensive description using the Vavilov and Landau distributions is also available using modified routines from GEANT3. A simple model is adopted for the multiple Coulomb scattering, based on a Gaussian distribution. The need to use fast multiple scattering sampling distributions is a major cause for uncertainty in AlfaMC at low energies.

When slowing down in matter ions will capture electrons from the medium reducing the effective electrical charge. The effect is somehow taken into account in the NIST/ASTAR stopping-power values since in the low-energy region they are calculated from fitting formulas that represent experimental data for many elements and a limited number of compounds. The effect is not taken into account in the computation of the energy straggling or multiple scattering, since a reduction in the effective charge from +2e would decrease both effects at low energy, increasing the difference to SRIM results.

The comparison with the well established SRIM program shows that AlfaMC can deliver meaningful results in the 1 to 12 MeV range for the alpha particle's energy. For energies lower than 1 MeV, AlfaMC results are less accurate and differences greater than 10% can be found between the results of both programs.

The AlfaMC program speed was compared with SRIM and the general-purpose Monte Carlo code (FLUKA) running on the same CPU for a simple slab geometry. A 10 MeV alpha particle beam was set to irradiate a 10 $\mu$m-thick Al foil. Under these conditions the particle losses about 1 MeV crossing the foil and eventual different cut-offs have little influence in program speed. This is important since SRIM does not allow any user cut-off control. It was found that FLUKA would take 5.5 more time than AlfaMC per primary particle, while SRIM would take 63 more time per event than AlfaMC.

The AlfaMC code, as well as the ULYSSES and ULHISTOS packages are open-source codes released under the General Public Licence (GPL) and can be obtained at http://www.lip.pt/alfamc . A package containing the necessary GEANT3 routines is also, included in the standard AlfaMC distribution file. These programs are also distributed under GPL.

**References**

Agostinelli, S. et al., 2003. Geant4-a simulation toolkit, Nuclear Instruments and Methods in Physics Research Section A: Accelerators, Spectrometers, Detectors and Associated Equipment, Volume 506, Issue 3, Pages 250-303

Allison, J. et al., 2006. Geant4 developments and applications, IEEE Transactions on Nuclear Science 53 No. 1, 270-278.

Atomic weights of the elements, 2009. IUPAC Technical Report,
 http://www.chem.qmul.ac.uk/iupac/AtWt/index.html (accessed 2012)

Attix, F.H., 2008. Introduction to Radiological Physics and Radiation Dosimetry, John Wiley & Sons

Berger, M.J., Coursey, J.S., Zucker, M.A. and Chang, J., 2011. Stopping-power and range tables for helium ions, http://physics.nist.gov/PhysRefData/Star/Text/ASTAR.html (accessed 2012)

Beringer, J. et al. (Particle Data Group), 2012. The Review of Particle Physics, Phys. Rev. D86, 010001, http://pdg.lbl.gov/, accessed 2012

Bethe, H.A., 1953. Phys. Rev. 89, 1256.

Bohr, N. 1913. On the theory of the decrease of velocity of moving electrified particles on passing through matter', *Philosophical Magazine 25*, *10*–31

Chu, W. K., 1976. Calculation of energy straggling for protons and helium ions , Phys. Rev. A 13, 2057-2060.

Davies, H. , Bethe, H.A. and Maximon, L.C. , 1954. Phys. Rev. 93, 788.

FLUKA, 2012. http://www.fluka.org (accessed 2012)

GEANT, 1993. CERN Program Library Long Writeup W5013, GEANT Detector Description and simulation tool, http://wwwinfo.cern.ch/asdoc/pdfdir/geant.pdf (accessed 2012)

GEANT4, 2011. http://geant4.web.cern.ch/geant4/UserDocumentation/UsersGuides/PhysicsReferenceManual/fo/PhysicsReferenceManual.pdf (accessed 2012)

Geary, M.J. and Haque, K. M. M., 1976. The stopping power and straggling for alpha particles in tissue equivalent materials, NIM 137, 151-155

Hatano, Y., Katsumura, Y. and Mozumder, 2011, Charged Particle and Photon Interactions with Matter, recent adavances, applications and interfaces, CRC Press, pag. 327

Highland, V.L. , 1975. Nucl. Instr. and Meth. 129, 497

Highland, V.L. , 1979. Nucl. Instr. and Meth. 161, 171.

ICRU, 1993. International Commission on Radiation Units and Measurements. *ICRU Report 49*, *Stopping Powers and Ranges for Protons and Alpha Particles*.

Johns, H.E. and Cunningham, J.R., 1983. The physics of radiology, ed. Charles C. Thomas.

Knoll, G.F. , 2010. Radiation Detection and Measurement , ed. John Wiley & Sons.

Landau, L.D. 1944, On the Energy Loss of Fast Particles by Ionization, Jour. Phys. USSR 8, 201

Leo, W.R., 1994. Techniques for Nuclear and Particle Physics Experiments: A How-To Approach, Springer

Livingston, M. S. and Bethe, H., 1937. Rev. Mod. Phys. 9, 245

Lynch, G.R. and Dahl, O.I, 1991. Approximations to multiple Coulomb scattering, Nucl. Instrum. Methods B58, 6.

MCNPX, 2012. http://mcnpx.lanl.gov/ (accessed 2012)

Nelson W.R. et al., 1985. The EGS4 code system. Technical Report 265, SLAC.

NIST, 2012a http://physics.nist.gov/PhysRefData/Star/Text/programs.html (accessed 2012)

NIST, 2012b http://physics.nist.gov/PhysRefData/XrayMassCoef/tab1.html (accessed 2012)

Payne, M. G., 1969. Energy Straggling of Heavy Charged Particles in Thick Absorbers , Phys. Rev. 185, 611-623

Ramirez, J.J. , Prior, R.M., Swint, J.B. , Quinton, A.R. and Blue, R. A., 1969. Energy Straggling of alpha particles through gases, Phys. Rev. 179, 310-314.

Rossi, B. and Greisen, K. , 1941. Rev. Mod. Phys. 13, 240.

Seltzer, S.M. and Berger, M.J. 1964. Energy loss straggling of protons and mesons. In Studies in Penetration of Charged Particles in Matter. Nuclear Science Series 39, Nat. Academy of Sciences, Washington DC.

Schorr, B., 1974. Programs for the Landau and the Vavilov distributions and the corresponding random numbers. Comp. Phys. Comm., 7, 216.

Strittmatter, R.B. and Wehring, B. W., 1976. Alpha-particle energy straggling in solids , Nucl. Instr. Methods 135 173-177

SRIM, 2008. http://www.srim.org/ (accessed 2012)

Table, 2004. R.B. Firestone and L.P. Ekström,Table of Radioactive Isotopes, Electronic edition, http://ie.lbl.gov/toi/ (acessed 2012)

Tsai, Y.S. , 1974. Rev. Mod. Phys. 46, 815.

C. Tschalar, C. and Maccabee, H. D., 1970. Energy-Straggling Measurements of Heavy Charged Particles in Thick Absorbers , Phys. Rev. B 1, 2863-2869.

Tsoulfanidis, N., 1995. Measurement And Detection Of Radiation, Taylor & Francis

Turner, J.E., 2007. Atoms, Radiation, and Radiation Protection, edited by Wiley-VCH

Ulysses, 2012. http://www.lip.pt/ulysses/Ulysses_manual.pdf (accessed 2012)

Ulhistos, 2012. http://www.lip.pt/ulysses/Ulhistos_manual.pdf (accessed 2012)

Vavilov, P.V., 1957. Ionisation losses of high energy heavy particles. Soviet Physics JETP, 5, 749-758.

Wieser, M.E. and Coplen, T.B., 2011. Pure Appl. Chem. Vol. 83, No. 2, pp. 359-396

http://dx.doi.org/10.1351/PAC-REP-10-09-14

Wong, M., Schimmerling, W. , Phillips, M.H. , Ludewigt, B.A., Landis, D.A., Walton, J.T., Curtis, S.B. , 1990. The multiple Coulomb scattering of heavy charged particles, Med. Phys. 17, 163-171

Ziegler, J. F.and Biersack, J. P., 1985, The Stopping and Range of Ions in Solids, Pergamon Press.

## Appendix A

**Rotation from the particle reference frame to the laboratory reference frame**

Let $\vec{u}_0 = (u, v, w)$ be the unit vector of the particle direction. The parameters (u,v,w) are the cosine of the projection angles on the x,y,z axis

$$u = \cos\alpha_x$$
$$v = \cos\alpha_y$$
$$w = \cos\alpha_z$$

We choose the particle reference frame to have the z axis in the particle's motion direction and thus $\vec{e}'_z = u\vec{e}_x + v\vec{e}_y + w\vec{e}_z$

We notice the choice of the x' and y' axis in this reference frame is arbitrary as long as the x',y',z' axis are orthogonal to each other. A possible choice is to define the x' axis in the particle reference frame as

$\vec{e}'_x = \vec{e}'_z \times \vec{e}_z / \|\vec{e}'_z \times \vec{e}_z\|$ and the external product is given by the following determinant

$$\vec{e}'_z \times \vec{e}_z = \begin{vmatrix} \vec{e}_x & \vec{e}_y & \vec{e}_z \\ u & v & w \\ 0 & 0 & 1 \end{vmatrix} = v\vec{e}_x - u\vec{e}_y \ .$$

We have for the norm $\|\vec{e}'_x\| = \|\vec{e}'_z \times \vec{e}_z\| = \sqrt{u^2 + v^2}$, so finally the unit vector of x' frame is

$$\vec{e}'_x = \frac{1}{\sqrt{u^2 + v^2}} (v\vec{e}_x - u\vec{e}_y) \ .$$

For the y axis unit vector we have $\vec{e}'_y = \vec{e}'_z \times \vec{e}'_x / \|\vec{e}'_z \times \vec{e}'_x\|$ , thus

$$\vec{e}\,'_z \times \vec{e}\,'_x = \begin{vmatrix} \vec{e}_x & \vec{e}_y & \vec{e}_z \\ u & v & w \\ \dfrac{v}{\sqrt{u^2+v^2}} & \dfrac{-u}{\sqrt{u^2+v^2}} & 0 \end{vmatrix} = \dfrac{1}{\sqrt{u^2+v^2}}(uw\,\vec{e}_x + vw\,\vec{e}_y - (u^2+v^2)\,\vec{e}_z)$$

It can be easily verified that $\|\vec{e}\,'_z \times \vec{e}\,'_x\| = 1$ .

We thus have

$$\vec{e}\,'_x = \dfrac{v}{D}\vec{e}_x - \dfrac{u}{D}\vec{e}_y + 0\,\vec{e}_z$$

$$\vec{e}\,'_y = \dfrac{uw}{D}\vec{e}_x + \dfrac{vw}{D}\vec{e}_y - \dfrac{(u^2+v^2)}{D}\vec{e}_z$$

$$\vec{e}\,'_z = u\,\vec{e}_x + v\,\vec{e}_y + w\,\vec{e}_z$$

where we defined $D = \sqrt{u^2+v^2}$ .

In matrix notation these relation provide the rotation from the laboratory frame to the particle frame

$$\begin{pmatrix} x' \\ y' \\ z' \end{pmatrix} = \begin{pmatrix} v/D & -u/D & 0 \\ uw/D & vw/D & -(u^2+v^2)/D \\ u & v & w \end{pmatrix} \begin{pmatrix} x \\ y \\ z \end{pmatrix}$$

The rotation from the particle frame to the laboratory frame is given by the inverse matrix, which is the transposed matrix

$$\begin{pmatrix} x \\ y \\ z \end{pmatrix} = \begin{pmatrix} v/D & uw/D & u \\ -u/D & vw/D & v \\ 0 & -(u^2+v^2)/D & w \end{pmatrix} \begin{pmatrix} x' \\ y' \\ z' \end{pmatrix} .$$

The choice made for the x' and y' breaks in the case the particle runs parallel to the z axis (i.e. $\vec{e}_z = \vec{e}\,'_z$ . In this case, both reference frames (laboratory and particle) are coincident.